\documentclass[aip,jap,reprint,floatfix]{revtex4-2}
\usepackage{graphicx}
\usepackage{amsmath,amssymb,bm}
\usepackage{siunitx}
\usepackage[hidelinks,colorlinks=true]{hyperref}
\usepackage{orcidlink}
\begin{document}

\title{Comparison of NiFeCr and NiFe in ferromagnetic Josephson junctions}
\author{Joshua Willard \orcidlink{0009-0007-8040-2315}}
\author{Swapna Sindhu Mishra \orcidlink{0000-0003-4074-4795}}
\author{Robert M. Klaes \orcidlink{0000-0002-1152-848X}}
\author{Nicholas J. Emtage \orcidlink{0000-0002-4570-6619}}
\author{Reza Loloee}
\author{Norman O. Birge \orcidlink{0000-0001-7840-3908}}
\email{birge@msu.edu}
\affiliation{Department of Physics and Astronomy, Michigan State University, East Lansing, MI 48824, USA}
\date{\today}

\begin{abstract}
Josephson junctions containing ferromagnetic materials are under consideration for applications in digital superconducting logic and memory.  Some memory applications rely on the ability to reverse the magnetization direction of a ``soft" magnetic layer within the junction using a small local magnetic field generated on the chip. It is crucial, therefore, to find a suitable soft magnetic material with a low switching field and low switching energy.  A popular magnetic material for such applications is Ni$_{80}$Fe$_{20}$, also known as Permalloy, however Permalloy has a rather large magnetization, leading to large magnetic switching energies.  In this work we explore Cr-doped Permalloy, specifically Ni$_{73}$Fe$_{18}$Cr$_{9}$, which has a saturation magnetization just under two-thirds that of Permalloy.  Josephson junctions containing this NiFeCr alloy undergo a 0-$\pi$ transition at a NiFeCr thickness of \SI{2.3}{\nano\meter}, and the critical supercurrent decays in the alloy over a short characteristic length of \SI{0.36}{\nano\meter}. Switching fields of a few millitesla are promising, but the short decay length and overall small values of the critical current in the Josephson junctions may preclude the use of NiFeCr in current cryogenic memory technologies.
\end{abstract}

\maketitle

\section{Introduction}
Josephson junctions containing ferromagnetic (F) materials have been subject to intense study since the discovery that the ground-state phase difference across the junction can be either 0 or $\pi$ depending on the thickness of the F layer(s).\cite{ryazanov_2001,kontos_2002}  The occurrence of such ``$\pi$-junctions" had been predicted 20 years earlier,\cite{bulaevskii_1978,buzdin_1982,buzdin_1991} but the fabrication technology of metallic multilayers was not mature enough at that time to realize $\pi$-junctions in the laboratory.  Nowadays $\pi$-junctions are fabricated in numerous laboratories around the world, including some industrial-scale superconducting circuit foundries.\cite{dayton_2018,tolpygo_2019}

Ferromagnetic $\pi$-junctions have potential uses in superconducting logic and memory,\cite{ustinov_2003,ortlepp_2006,khabipov_2010,ryazanov_2012,soloviev_2017,kamiya_2018,takeshita_2021} and possibly even in quantum computing.\cite{ioffe_1999,blatter_2001,yamashita_2005,feofanov_2010} Here we focus on applications in classical superconducting memory, where some property of the junction -- either critical current or phase -- can be modulated by changing the magnetic configuration inside the junction.\cite{krivoruchko_2001,golubov_2002,bell_2004,baek_2014,abdelqader_2014,baek_2015,gingrich_2016,niedzielski_2018,dayton_2018,birge_2018,madden_2019,glick_2018} For such applications to be useful in a large-scale superconducting digital circuit, it is essential to minimize both the amplitude of the magnetic field needed to modify the magnetic state and the energy dissipated during magnetic switching. In the case of thin elliptical nanomagnets for which magnetization reversal occurs via coherent rotation (Stoner-Wohlfarth switching),\cite{stoner_1948} the field required to switch the magnetization direction is proportional to both the saturation magnetization and the thickness of the nanomagnet, $H_{\mathrm{sw}} \propto M_s t_F$, while the switching energy is proportional to the square of that product: $E_{\mathrm{sw}} \propto M_s^2 t_F^2$.\cite{ohandley_2000} Those relations suggest using weakly-ferromagnetic materials with low $M_s$, as long as the thickness $t_F$ needed to achieve a $\pi$-junction does not increase by the same factor that $M_s$ decreased. 

While the properties of many ferromagnetic materials have been well studied at room temperature,\cite{bozorth_1951} there are very few studies of such materials at cryogenic temperatures, and fewer still that explore both the magnetic properties and the capability of the material to carry supercurrent when inserted into a Josephson junction. That situation has started to improve over the past two decades.  One can now find published works reporting the behavior of Josephson junctions containing both pure elemental F materials as well as several of their alloys covering a wide range of $M_s$ values: Ni,\cite{blum_2002,shelukhin_2006,robinson_2006,bannykh_2009,baek_2018} Co,\cite{robinson_2006} Fe,\cite{robinson_2006} Ni$_3$Al,\cite{born_2006} NiFe,\cite{robinson_2006,abdelqader_2014,glick_2017,mishra_2022} NiFeNb,\cite{baek_2014,niedzielski_2014} NiFeCu,\cite{abdelqader_2014} NiFeMo,\cite{niedzielski_2015} NiFeCo,\cite{glick_2017} CuNi,\cite{ryazanov_2001,sellier_2003,oboznov_2006,weides_2006} PdNi,\cite{kontos_2002,khaire_2009,pham_2022} PdFe,\cite{larkin_2012,vernik_2013,caruso_2018,glick_2017a}  and most recently CoB.\cite{satchell_2020,satchell_2021}  (For a review, see Ref. \onlinecite{birge_2024}).  Based on these works we can draw some tentative conclusions about which materials are or are not likely to be useful in real-world applications. For example, Ryazanov and co-workers\cite{oboznov_2006} discovered that the critical current, $I_c$, in Josephson junctions containing CuNi alloy is rather small and decreases rapidly with the CuNi thickness, possibly due to spin-flip scattering by Ni clusters. As a result, CuNi is not an attractive material for use in circuits that require Josephson junctions with large critical currents.

Another weakly-ferromagnetic system that looks attractive at first sight are the doped Pd alloys: PdNi and PdFe.  PdNi was used by Kontos \textit{et al.} in their pioneering experiments on tunneling\cite{kontos_2001} and $\pi$-junctions.\cite{kontos_2002}  While PdNi is indeed an excellent material for making fixed $\pi$-junctions, it is less useful for memory applications that require magnetization switching.  PdNi has rather high coercivity and also has perpendicular magnetization anisotropy, which is inconvenient in situations where an applied magnetic field is used to switch the magnetization direction of a material that is sandwiched inside a Josephson junction.  Another option is PdFe, which has several advantages.  PdFe is magnetic even with Fe concentrations below 0.1\%,\cite{bscher_1992} hence its magnetization and Curie temperature can be tuned over wide ranges with only a small variation in Fe concentration. Due to the very low spin-orbit coupling of Fe impurities in Pd,\cite{khoi_1976,senoussi_1977,arham_2009} there is almost no magnetocrystalline anisotropy in very dilute PdFe alloys, hence one would hope that the switching characteristics would be completely determined by the engineered shape anisotropy of the memory element. The possibility of using dilute PdFe in a memory element was proposed in a series of papers from the Hypres-Chernogolovka collaboration.\cite{larkin_2012,ryazanov_2012,vernik_2013} Our own experience with PdFe, however, has been less than satisfactory. While it is easy to control the PdFe thickness to fabricate a $\pi$-junction, careful analysis of the critical current vs applied field showed that the magnetization switching is a continuous process that starts close to zero field and finishes at a few tens of Oersted, suggesting that magnetization reversal occurs by pinned domain wall motion rather than by coherent rotation.\cite{glick_2017} That type of reversal is not optimal for a memory element.  Our conclusion from those results was that extremely weak ferromagnetic alloys are not the best choice if their exchange stiffness is too small to overcome magnetization pinning by defects and surface roughness that inevitably arise in nanoscale magnetic elements grown on top of a thick superconducting electrode.

The considerations discussed above led us back to Permalloy, a NiFe alloy with approximate concentration Ni$_{80}$Fe$_{20}$.  NiFe has a rather large saturation magnetization, $M_s \approx$ \SI{900}{\kilo\ampere\per\meter} at low temperature,\cite{ohandley_2000} but it can be doped with a large number of nonmagnetic elements to reduce $M_s$.\cite{bozorth_1951}  Doping comes with a warning, however. For example, doping with Mo should produce a very soft magnetic material (Supermalloy) suitable for applications at low temperature.  But Josephson junctions containing NiFeMo exhibit very small values of $I_c$ and a rapid decrease of $I_c$ with thickness,\cite{niedzielski_2015} -- similar to what was observed in CuNi alloy.\cite{oboznov_2006} Another candidate, NiFeNb, can be made to be very soft with excellent switching characteristics at low temperature,\cite{baek_2014} but it severely depresses the $I_c$ of the Josephson junctions in which it is placed. Qadar \textit{et al.} performed a thorough study of the structural and magnetic properties of NiFeCu spanning a wide range of Cu concentrations.\cite{qader_2017}  A disappointing result of that study was that the coercive field of thin NiFeCu films stayed above 100 Oe for Cu concentrations ranging between 20\% and 85\%, suggesting that the magnetic properties of the films with high Cu concentration are strongly affected by disorder. While high coercivity due to domain wall pinning does not preclude magnetization reversal by coherent rotation in sufficiently small magnetic structures, the appearance of a large coercive field in soft magnetic materials is generally not a good sign as it suggests that the magnetic properties are strongly affected by extrinsic effects such as defects, surface roughness, or magnetic inhomogeneity. 

The results on NiFeCu discussed above led us to consider NiFeCr.  The magnetic moments of Cr impurities in NiFe point opposite to the Ni and Fe moments,\cite{bozorth_1951} hence one should be able to reduce $M_s$ in NiFeCr using far less Cr than what is required for a similar reduction of $M_s$ in NiFeCu. Devonport \textit{et al.} have studied the structural and magnetic properties of NiFeCr films with Cr concentrations ranging from 0 to 35\%.\cite{devonport_2018}  Those authors find that the magnetic properties of the films at both room temperature and cryogenic temperature (\SI{10}{\kelvin}) remain excellent for Cr concentration up to about 15\%, at which point $M_s$ has fallen to about half its value in NiFe.  Those promising results provided the main motivation for the work presented here. In this work we chose to study an alloy with 9\% Cr to ensure that the Curie temperature remained above room temperature.

\section{Methods}

\subsection{Thin films for magnetic characterization}
Nb(5)/Cu(2)/NiFeCr($d_{\mathrm{NiFeCr}}$)/Cu(2)/Nb(5) (layer thicknesses are in nanometers) thin films were grown on an Si/SiO$_2$ substrate with dc magnetron sputtering. Energy dispersive X-ray (EDX) analysis of thicker films showed compositions of Ni$_{73}$Fe$_{18}$Cr$_{9}$ and Ni$_{82}$Fe$_{18}$ for NiFeCr and NiFe, respectively. The base pressure of the sputtering chamber before the deposition was \SI{5E-6}{\pascal} and the deposition process was performed at an Ar pressure of \SI{0.3}{\pascal} and a substrate temperature around \SI{250}{\kelvin}. $d_{\mathrm{NiFeCr}}$ was varied from \SI{1}{} to \SI{4.5}{\nano\meter} in steps of \SI{0.5}{\nano\meter}.  We place a small magnet behind our samples during the sputtering process to orient the magnetocrystalline anisotropy of the NiFeCr and NiFe films in the desired direction. 

The magnetic moment versus magnetic field measurements for all thin film samples were performed at a temperature of \SI{10}{\kelvin} using Quantum Design MPMS3 which is a SQUID-based vibrating sample magnetometer.

\subsection{Josephson junctions}
We discuss our Josephson junction fabrication process here briefly but previously published works contain a more detailed description.\cite{glick_2017} First, the bottom lead stencil was patterned on a clean Si/SiO$_2$ substrate using a masked photolithography process. Then [Nb(25)/Al(2.4)]$_3$/Nb(20)/Cu(2)/NiFeCr($d_{\mathrm{NiFeCr}}$) /Cu(2)/Nb(5)/Au(10) was sputtered where $d_{\mathrm{NiFeCr}}$ was varied from \SI{1.0}{} to \SI{3.7}{\nano\meter}. After a liftoff process, we obtain our bottom leads. The Josephson junctions were then patterned by e-beam lithography, followed by ion-milling to form the pillars and $\mathrm{SiO_x}$ deposition \textit{in-situ} around the junction area to avoid electrical shorts between the bottom and top superconducting electrodes to be deposited next. The junction shape is elliptic cylindrical with major and minor axis nominal dimensions of \SI{1.25}{\micro \meter} and \SI{0.5}{\micro \meter}, respectively. The major axis is oriented along the magnetocrystalline easy axis set during the bottom lead deposition. After liftoff, the top lead stencil was patterned using masked photolithography. Roughly half of the Au(10) capping layer from the bottom lead deposition was ion milled \textit{in-situ} to improve surface contact and then Nb(150)/Au(10) layers were deposited by sputtering. Top superconducting electrodes are thus formed after a subsequent liftoff process and the junctions are ready for measurement.

The Josephson junctions are mounted on a home-built probe with a built-in superconducting magnet and connected to four-probe measurement setup. The probe is then inserted inside a liquid He$^4$ dewar for electrical transport measurements at \SI{4.2}{\kelvin}. $I-V$ curves for all Josephson junctions were measured in magnetic fields up to fields of \SI{80}{\milli\tesla} in both directions. The magnetic field $H$ was directed along the major axes of the elliptical junctions in the plane of the sample. Most of the junctions were measured using standard commercial electronics: a current source and nanovoltmeter.  A few of the junctions containing NiFeCr had very low values of $I_c$; junctions with maximum critical currents less than \SI{2}{\micro\ampere} were measured using a battery-powered ultra-low-noise current source and an rf-SQUID-based self-balancing potentiometer circuit with voltage noise of a few $\SI{}{\pico\volt}/\sqrt{\textrm{Hz}}$.\cite{glick_2017}    

\section{Results}
\subsection{Thin film magnetics}
\begin{figure}[!htbp]
\includegraphics[width=\linewidth]{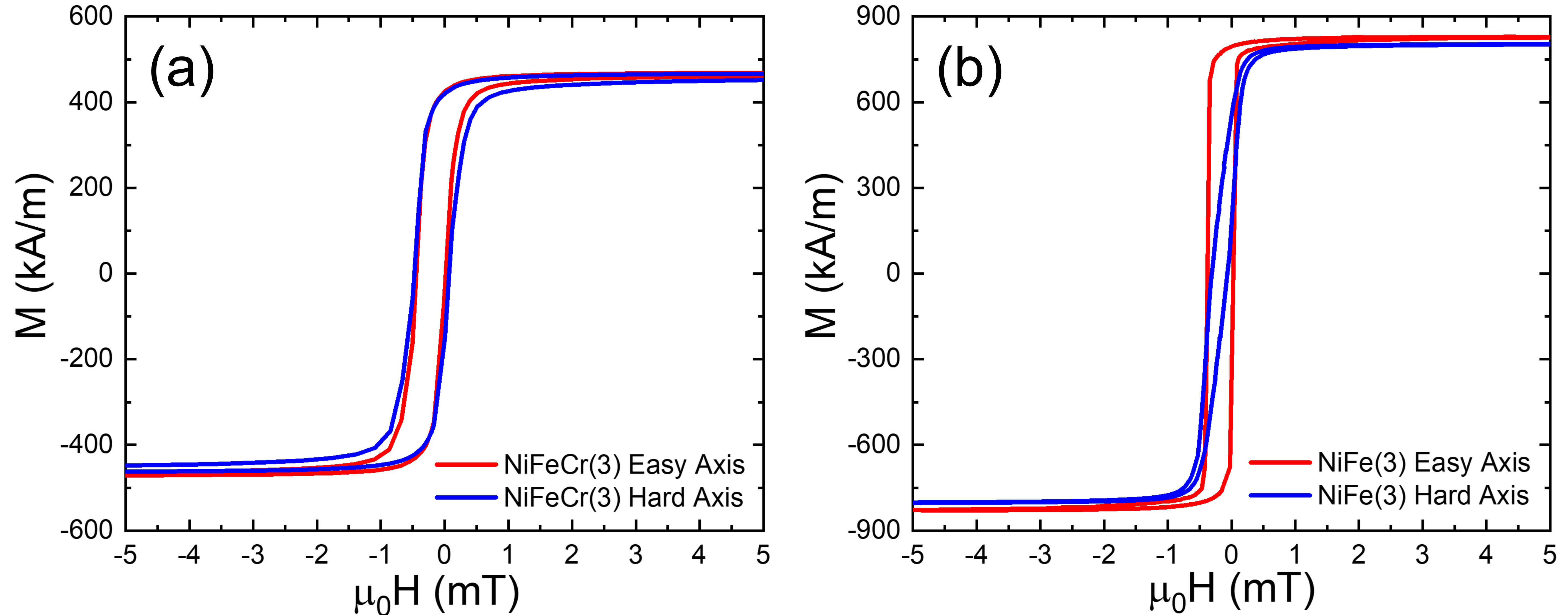}
\caption{Magnetization ($M$) vs field ($H$) measured at $T = \SI{10}{\kelvin}$ for (a) NiFeCr(3) and (b) NiFe(3) films with the field aligned along the easy (red) and hard (blue) axis.}
\label{fig:MvH}
\centering
\end{figure}

The magnetization ($M$) versus field ($H$) for NiFeCr(3) and NiFe(3) thin films measured along both the easy and hard axes at \SI{10}{\kelvin} are shown in Fig \ref{fig:MvH}.  The NiFe sample shows a clear magnetocrystalline anisotropy; the easy axis data exhibit close to 100\% remanance while the hard-axis data exhibit very small remanence.  The NiFeCr sample, on the other hand, exhibits very little magnetocrystalline anisotropy. That is not necessarily a problem for memory applications; as long as the coercivity is low, then the anisotropy of a memory bit will be determined largely by its shape.

\begin{figure}[!htbp]
\includegraphics[width=\linewidth]{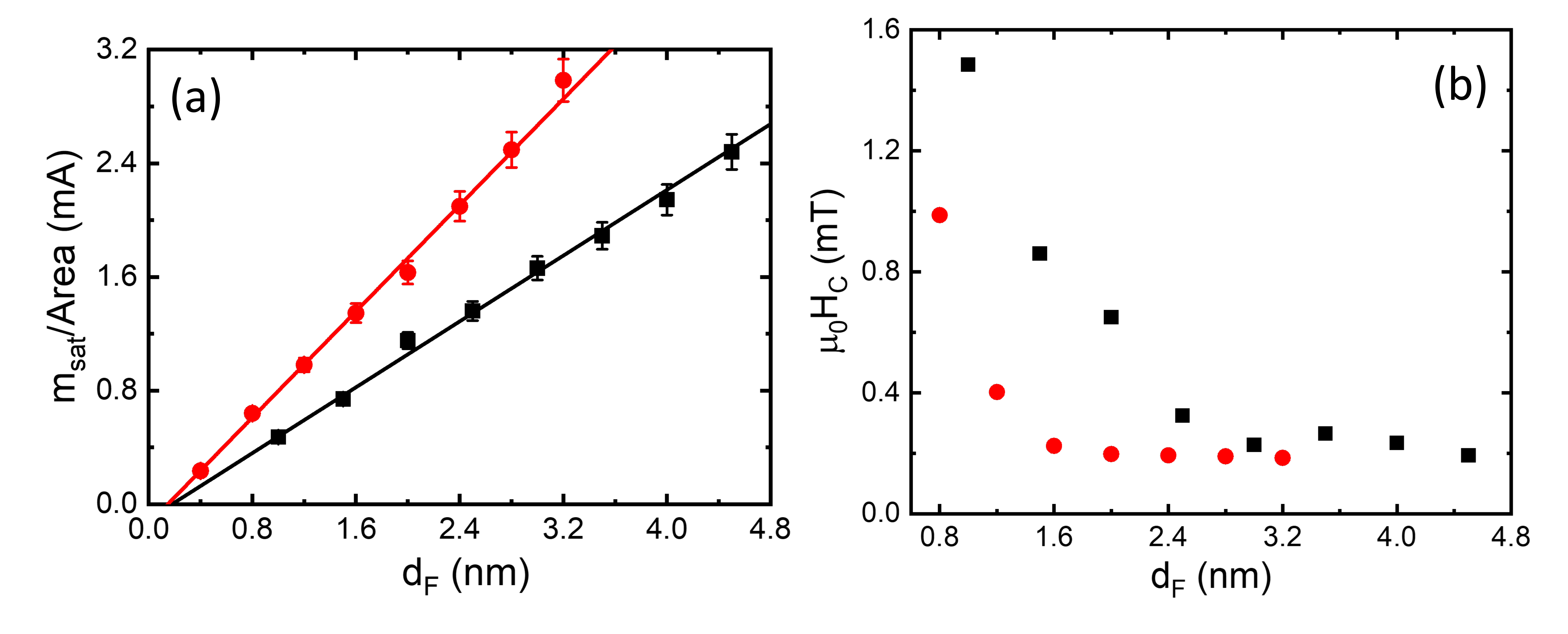}
\caption{(a) Saturation moment per unit area ($m_{\mathrm{sat}}/\mathrm{Area}$ and (b) Coercivity ($H_c$) vs thickness ($d_{\mathrm{F}}$) for NiFeCr($d_{\mathrm{NiFeCr}}$) (black squares) and NiFe($d_{\mathrm{NiFe}}$) (red circles) films, measured at $T = \SI{10}{\kelvin}$. The solid lines are linear fits.}
\label{fig:MsatHc}
\centering
\end{figure}

The saturation moment per unit area versus thickness for NiFeCr($d_{\mathrm{NiFeCr}}$) and NiFe($d_{\mathrm{NiFe}}$) samples is shown in Fig. \ref{fig:MsatHc}(a). Since area estimation is made using an optical microscope, an error of 5\% is assigned to each data point. Values of the magnetizations are determined from the slopes of the linear fits to be \SI[separate-uncertainty=true, multi-part-units=single]{579 \pm 16}{\kilo\ampere\per\meter} for NiFeCr and \SI[separate-uncertainty=true, multi-part-units=single]{935 \pm 17}{\kilo\ampere\per\meter} for NiFe. From the intercepts, we estimate the combined dead layer thicknesses of the two interfaces with Cu to be \SI[separate-uncertainty=true, multi-part-units=single]{0.18 \pm 0.05}{\nano\meter} and \SI[separate-uncertainty=true, multi-part-units=single]{0.15 \pm 0.01}{\nano\meter} for NiFeCr and NiFe, respectively. 

The coercivity versus thickness for NiFeCr($d_{\mathrm{NiFeCr}}$) and NiFe($d_{\mathrm{NiFe}}$) samples is shown in Fig. \ref{fig:MsatHc}(b). To eliminate the effect of the small field shift due to trapped flux in the magnet, the coercive fields for each field sweep direction were determined by fitting the raw M vs H curves to an error function; the plotted coercive field is the average of the two absolute values. For the thickest films, the coercivity of NiFeCr is comparable to that of NiFe, but the coercivity grows considerably faster in NiFeCr than in NiFe as the thickness decreases.  That may be due to pinning by defects and surface roughness along with increased magnetostriction in the NiFeCr.

\subsection{Josephson junction transport}
Josephson junctions that contain ferromagnetic materials without an insulating barrier exhibit overdamped dynamics. In such cases, the current-voltage ($I-V$) curves follow the Resistively Shunted Junction model:\cite{barone_1982}
\begin{equation}
    V = \mathrm{sign}(I) R_N \Re\left\{\sqrt{I^2-I_c^2}\right\}
\end{equation}
where $I_c$ is the critical current, $R_N$ is the normal-state resistance of the junction and $\Re$ represents the real part of the argument. Values of $I_c$ and $R_N$ are obtained from a fit of the above equation to the experimental data.  As mentioned earlier, a few of the NiFeCr junctions had very low values of $I_c$; in such cases the $I-V$ data exhibit substantial rounding due to thermal and environmental noise. For the samples with maximum values of $I_c$ less then \SI{2}{\micro\ampere}, we fit the raw $I-V$ data with the Ivanchenko-Zil'berman function, with an effective noise temperature of about \SI{13}{\kelvin}.\cite{ivanchenko_1969,ambegaokar_1969}

\begin{figure}[!htbp]
\includegraphics[width=\linewidth]{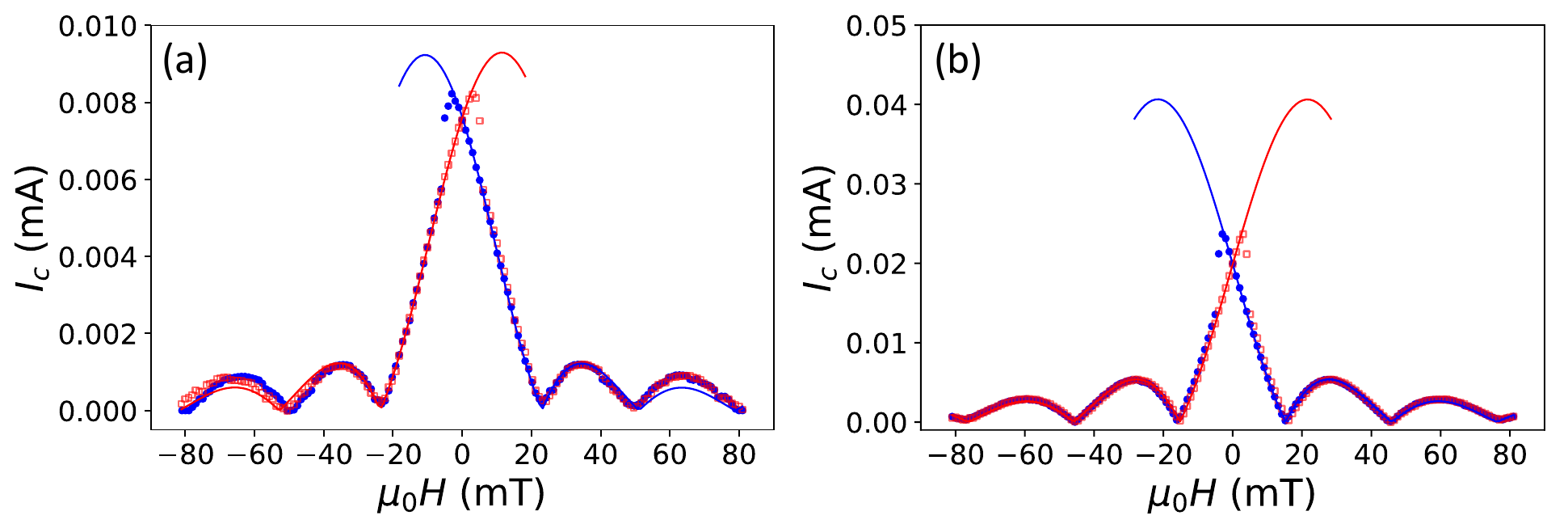}
\caption{Critical current ($I_c$) vs field ($H$) for (a) NiFeCr(2.9) and (b) NiFe(3.7). The blue circles and red squares represent the data taken during downsweep and upsweep of the magnetic field, respectively. The blue and red solid lines are fits to Eqn. \ref{Eqn:Airy} for downsweep and upsweep, respectively. These samples are representative of all the samples in the respective data sets.}
\label{fig:Fraunhofers}
\centering
\end{figure}

The dependence of $I_c$ on the applied magnetic field $H$ for two representative Josephson junctions containing NiFeCr(2.9) and NiFe(3.7) are shown in Fig. \ref{fig:Fraunhofers}(a) and \ref{fig:Fraunhofers}(b), respectively. The blue and red data points were acquired during the field downsweep and upsweep, respectively.

For elliptically shaped junctions, the experimental data are expected to follow an Airy function when the field is applied along a principal axis:\cite{barone_1982} 
\begin{equation} \label{Eqn:Airy}
    I_c(\Phi) = I_{c0} \left| \frac{2 J_1 \left( \frac{\pi \Phi}{\Phi_0} \right) }{\frac{\pi \Phi}{\Phi_0}} \right|
\end{equation}
where $I_{c0}$ is the maximum value of $I_c$, $J_1$ is the Bessel function of the first kind, and $\Phi_0 = 2.07 \times 10^{-15}$ Tm$^2$ is the flux quantum.  If the magnetization $M$ is uniform inside the junction, then the total magnetic flux through the junction is given approximately by:
\begin{equation} \label{Eqn:Flux}
    \Phi = \mu_0 H w (2\lambda_{\mathrm{eff}} + d_N + d_F) + \mu_0 M w d_F
\end{equation}
where $\lambda_{\mathrm{eff}}$ is the effective London penetration depth of the superconducting electrodes, $d_N$ is the thickness of any normal (non-ferromagnet/non-superconductor) layers, $d_F$ is the thickness of the ferromagnetic layer, and $w$ is the width of the junction transverse to the field direction. The solid lines in Fig. \ref{fig:Fraunhofers} are fits of Eqns. (\ref{Eqn:Airy}) and (\ref{Eqn:Flux}) to the experimental data. Each field sweep was fit separately to the same functions. The center of the Airy pattern exhibits a hysteretic shift in either direction due to the internal magnetization of the ferromagnetic layer in the junction. $I_c$ exhibits discontinuous drops shortly after $H$ passes through zero when the magnetization of the F layer switches direction. Because of that switch, the value of $I_{c0}$ extracted from the fits to Eqn. (\ref{Eqn:Airy}) is usually higher than the maximum measured value. 

\begin{figure*}[!htbp]
\includegraphics[width=\linewidth]{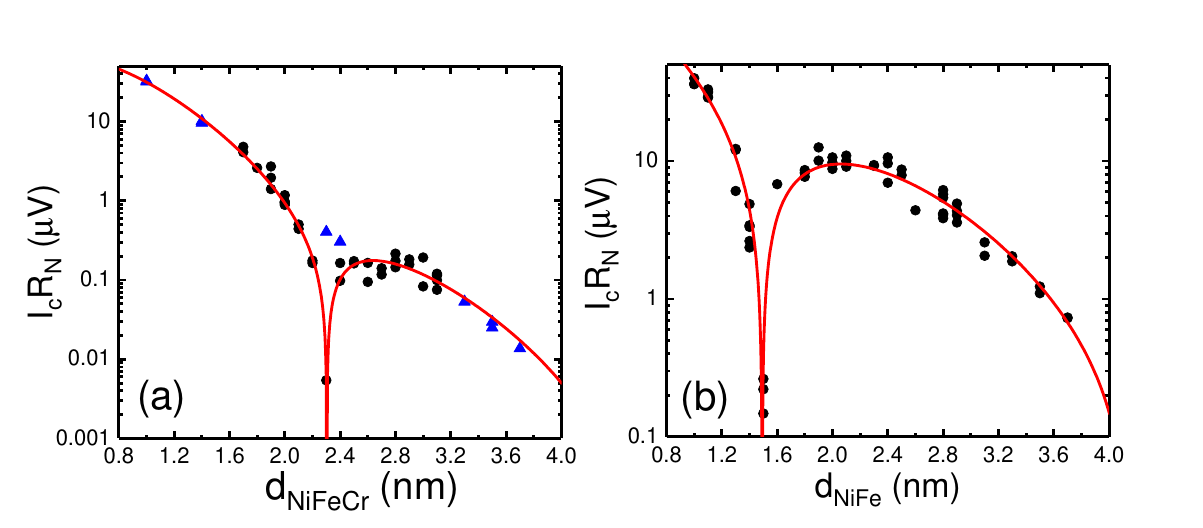}
\caption{$I_c R_N$ vs ferromagnetic layer thickness for Josephson junctions containing (a) NiFeCr($d_{\mathrm{NiFeCr}}$) and (b) NiFe($d_{\mathrm{NiFe}}$). The solid red lines are fits of the data to Eqn. (\ref{Eqn:Diffusive}). The black circles and blue triangles in panel (a) represent junctions fabricated in different sputtering runs. The figure in panel (b) was published previously in Ref. \onlinecite{mishra_2022}.}
\label{fig:ICRN}
\centering
\end{figure*}

Since $I_{c0}$ is proportional to the junction area, we multiply it by the normal state resistance $R_N$ to obtain the value of $I_c R_N$ for each sample. $I_c R_N$ is independent of variations in junction area that may arise from fabrication inconsistencies. Figure \ref{fig:ICRN} shows $I_c R_N$ versus F-layer thickness both for Josephson junctions containing NiFeCr (a), and for comparison, for junctions containing NiFe (b). Both sets of junctions exhibit a deep minimum in $I_c R_N$, indicating the transition between $0$ and $\pi$ junctions.

\section{Discussion}

The dependence of $I_c R_N$ on the ferromagnet thickness has been calculated theoretically in several different limits and measured experimentally by many groups for different ferromagnetic materials.\cite{birge_2024} The behavior of $I_c R_N$ versus ferromagnet thickness is predicted to oscillate and decay, either algebraically for ballistic transport \cite{buzdin_1982} or exponentially for diffusive transport.\cite{buzdin_1991} In the case of weak ferromagnets, where majority and minority spin bands are nearly identical, this oscillatory-decay behavior can be calculated in the diffusive limit using the Usadel equations.\cite{buzdin_1991} In the absence of spin-flip or spin-orbit scattering, both the oscillation and decay are governed by the diffusive-limit ``ferromagnetic coherence length" or ``exchange length", $\xi_F^*=\sqrt{\hbar D_F/2E_{\mathrm{ex}}}$, where $E_{\mathrm{ex}}$ and $D_F$ are the exchange energy and diffusion constant in F. However in the case of strong ferromagnets with a large exchange energy separating the majority and minority spin bands, those equations are not strictly valid. A formula more appropriate for strong magnetic materials was derived by Bergeret \textit{et al.} using the Eilenberger equation;\cite{bergeret_2001a}  their ``intermediate limit" formula also predicts an exponentially-decaying, oscillating function, but with the decay length given by the mean free path rather than the exchange length, and the oscillation length given by the clean-limit expression for the exchange length, $\xi_F=\hbar v_F/2E_{\mathrm{ex}}$, where $v_F$ is the Fermi velocity in F. Since NiFeCr is a strong F material with relatively short mean free paths for both majority and minority band electrons,\cite{devonport_2018} it is not clear that our NiFeCr junctions fall into either of the two regimes described above.  Fortunately, both the diffusive limit and intermediate limit formulas can be modeled approximately by the following exponentially-decaying oscillatory function:
\begin{equation} \label{Eqn:Diffusive}
    I_c R_N = V_0\; \mathrm{exp}\left(\frac{-d_F}{\xi_{F1}}\right) \left| \mathrm{sin} \left( \frac{d_F - d_{0-\pi}}{\xi_{F2}} \right)  \right|
\end{equation}
where $V_0$ is a fictitious magnitude of $I_c R_N$ extrapolated to zero F-layer thickness, $\xi_{F1}$ and $\xi_{F2}$ are the length scales that control the decay and oscillation period in the ferromagnet F, and $d_{0-\pi}$ is the thickness where the first $0-\pi$ transition occurs. The solid lines in Fig. \ref{fig:ICRN} are fits of Eqn. (\ref{Eqn:Diffusive}) to the data with experimental uncertainties obtained from the Airy function fits. The uncertainties are smaller than the symbol size in Fig. \ref{fig:ICRN} and not visible. Overall, the fit describes the data very well, with the exception of a couple of NiFeCr junctions fabricated in a second sputtering run with thicknesses very close to the $0-\pi$ transition. (The purpose of the second run was to extend the NiFeCr thickness range at both small and large thicknesses.)

The fit parameters for both the NiFeCr and NiFe data sets are tabulated in Table \ref{tab:parameters}.  We also include parameters obtained from our old data set on junctions containing NiFeMo.\cite{niedzielski_2015} We note that the data in the vicinity of the $0-\pi$ transition for the NiFeCr junctions are remarkably similar to the data in the same range for the NiFeMo junctions.

\begin{table}[!htbp]
\caption{Parameter values determined from fitting Eqn. (\ref{Eqn:Diffusive}) to the data shown in Fig. \ref{fig:ICRN} for Josephson junctions containing Ni$_{73}$Fe$_{18}$Cr$_{9}$, Ni$_{82}$Fe$_{18}$, and Ni$_{73}$Fe$_{21}$Mo$_{6}$.  Fit parameters for the latter are taken from Ref. [\onlinecite{niedzielski_2015}]. The thickness range in the NiFeMo study did not extend far enough to allow a reliable estimate of $\xi_{F2}$, so we leave that entry blank.}
\label{tab:parameters}
    \centering
    \begin{tabular}{|c|c|c|c|c|}
\hline
F material & $V_0$ (\SI{}{\micro \volt}) & $\xi_{F1}$ (\SI{}{\nano\meter}) & $\xi_{F2}$ (\SI{}{\nano\meter}) & $d_{0-\pi}$ (\SI{}{\nano\meter})\\
 \hline 
NiFeCr & $527\pm42$ & $0.36\pm0.01$ & $0.67\pm0.03$ & $2.30\pm0.01$\\
NiFe & $329\pm48$ & $0.67\pm0.03$ & $0.85\pm0.03$ & $1.49\pm0.01$\\ 
NiFeMo & $150\pm50$ & $0.48\pm0.04$ & N.A. & $2.25\pm0.02$\\
 \hline
    \end{tabular}
\end{table}

Independent of the fits shown in Fig. \ref{fig:ICRN}, we make several observations from the data.  First, the $0-\pi$ transition in the NiFeCr junctions occurs at NiFeCr thickness of \SI{2.3}{\nano\meter}, as compared to \SI{1.5}{\nano\meter} in the NiFe junctions.  Second, $I_c$ decays much more quickly with thickness in NiFeCr compared to NiFe.  And third, the maximum value of $I_c$ in the $\pi$-state is nearly two orders of magnitude smaller in the NiFeCr junctions than in the NiFe junctions.  Clearly, NiFeCr would not be appropriate for applications demanding $\pi$-junctions with large critical current density. 

How should one interpret the parameter values shown in Table I? We begin with the caveat that additional data to map out the second 0-$\pi$ transition would be needed to be more confident in the fit value for $\xi_{F2}$. But, assuming the observed trend is born out, we argue that it's not unreasonable to explain the shift in the 0-$\pi$ transition from the reduced value of the the exchange energy, $E_{ex}$, in NiFeCr. However, this argument contains more nuance than it might initially seem. Assuming the diffusive limit theoretical expression for $\xi_F$, the reduction in $E_{ex}$ may be more than compensated by a reduction in $D_F$. Indeed, the resistivity of a thick NiFeCr film was measured to be \SI{880}{\nano\ohm\meter}, more than 7 times higher than the value of \SI{120}{\nano\ohm\meter} we obtained for a thick NiFe film, which implies small values of the mean free path and diffusion constant. That high resistivity for NiFeCr is consistent with previous results.\cite{rice_1976} However, it has been suggested that while the addition of Cr to NiFe causes a rapid decrease in the majority band mean free path, it hardly affects the already-short minority band mean free path.\cite{devonport_2018} Therefore, if it is the smaller minority band mean free path that determines the value of $D_F$ in the expression for $\xi_F$ in both metals, then the difference between $\xi_F^*$ for NiFe and NiFeCr would be due only to the reduced value of $E_{\mathrm{ex}}$ in the latter.

Regarding the very short value of $\xi_{F1}$ in the NiFeCr junctions relative to the NiFe junctions, we attribute that either to spin-flip scattering or to the much larger resistivity, while acknowledging that we don't know which mean free path -- majority or minority spin species -- is the limiting factor.  

\section{Conclusion}

In conclusion, doping NiFe with Cr appeared initially to be a promising route to lowering the magnetic switching energy of controllable Josephson junctions for applications in cryogenic memory. Cr doping with only 9\% reduces the magnetization significantly without raising the coercivity excessively.  Unfortunately, the addition of 9\% Cr to NiFe suppresses the Josephson junction critical current by nearly two orders of magnitude in the $\pi$ state.  In addition, the NiFeCr thickness required to achieve the $\pi$ state increased almost inversely proportionately to the decrease in magnetization, so that the switching energy is hardly changed. The search for an optimal soft magnetic material for cryogenic memory applications continues!

\begin{acknowledgments}
We thank V. Aguilar, T.F. Ambrose, J. Kingsley, M.G. Loving, A.E Madden, D.L. Miller and N.D. Rizzo for helpful discussions, and R. Dolleman for showing us an efficient numerical implementation of the Ivanchenko-Zil'berman function. We appreciate the technical assistance provided by D. Edmunds and B. Bi, and acknowledge the use of the W. M. Keck Microfabrication Facility at Michigan State University. This research was supported by Northrop Grumman Corporation.
\end{acknowledgments}

\section*{Author Declarations}
\subsection*{Conflict of Interest}
The authors have no conflicts to disclose.

\section*{Data Availability}
The data that support the findings of this study are available from the corresponding author upon reasonable request. 

\bibliography{References}

\end{document}